%% file: main.tex
\documentclass[twocolumn]{aastex631}  %%% ApJ

\input{pacchetti}

\input{journals}

\input{definitions}
\graphicspath{{figures/}}

\shorttitle{Bursty star formation at high-$z$}
\shortauthors{Gelli, Mason \& Hayward}

\begin{document}

\title{The impact of mass-dependent stochasticity at cosmic dawn}

\correspondingauthor{Viola Gelli}
\email{viola.gelli@nbi.ku.dk}

\author{Viola Gelli}
\author{Charlotte Mason}
\affiliation{Cosmic Dawn Center (DAWN)}
\affiliation{Niels Bohr Institute, University of Copenhagen, Jagtvej 128, 2200 København N, Denmark}
\author{Christopher C. Hayward}
\affiliation{Center for Computational Astrophysics, Flatiron Institute, 162 Fifth Avenue, New York, NY 10010, USA}

\begin{abstract}
JWST is unveiling a surprising lack of evolution in the number densities of ultraviolet-selected (UV) galaxies at redshift $z\gtrsim10$.
At the same time, observations and simulations are providing evidence for highly bursty star formation in high-$z$ galaxies, resulting in significant scatter in their UV luminosities. 
Galaxies in low-mass dark matter halos are expected to experience most stochasticity due to their shallow potential wells.
Here, we explore the impact of a mass-dependent stochasticity using a simple analytical model. We assume that scatter in the $\muv-M_h$ relation increases towards lower halo masses, following the decrease in halo escape velocity, $\sigma_\mathrm{UV} \sim M_h^{-1/3}$, independent of redshift.
Since low-mass halos are more dominant in the early universe, this model naturally predicts an increase in UV luminosity functions (LFs) at high redshifts compared to models without scatter. 
We make predictions for additional observables which would be affected by stochasticity and could be used to constrain its amplitude, finding:
(i) galaxies are less clustered compared to the no-scatter scenario, with the difference increasing at higher-$z$;
(ii) assuming star-bursting galaxies dominate the ionizing photon budget implies reionization starts earlier and is more gradual compared to the no-scatter case,
(iii) at fixed UV magnitude galaxies should exhibit wide ranges of UV slopes, nebular emission line strengths and Balmer breaks.
Comparing to observations, the mass-dependent stochasticity model successfully reproduces the observed LFs up to $z\sim12$. However, the model cannot match the observed $z\sim14$ LFs, implying additional physical processes enhance star formation efficiency in the earliest galaxies.
\end{abstract}


\section{Introduction}

During its first two years of observations, the James Webb Space Telescope (JWST) has revealed a puzzling abundance of luminous galaxies at high redshifts ($z>10$) that are challenging our understanding of early star formation.
Numerous observations of ultraviolet (UV) bright galaxies have been reported, and the photometrically derived UV luminosity functions (LF) show a surprising lack of evolution, across a range of luminosities, during the first $\sim500~\rm Myr$ after the Big Bang \citep[e.g.][]{Adams23, Atek23, Castellano22, Donnan24, Finkelstein23, Harikane23, McLeod24, PerezGonzalez23, Bouwens23}.
Most pre-JWST models struggle to reproduce the observed trends, prompting the emergence of numerous new theories in an effort to solve this discrepancy.

One possibility is a systematic enhancement of the median UV flux of high-$z$ galaxies. This enhancement may be achieved through different mechanisms, such as: higher star formation efficiencies \citep[][]{Mason23, Dekel23, Inayoshi22, Li23}, top-heavier initial mass functions \citep[although likely not sufficient to reproduce the observations up to the highest $z$, e.g.,][]{Rasmussen23, Trinca23}, or efficient dust removal \citep{Ferrara23, Fiore23}. 

The presence of active galactic nuclei (AGN) has also been explored as a possible source contributing to the enhancement of the UV luminosity of high-$z$ systems \citep[e.g.][]{Hegde24}; however, most of the JWST detected high-$z$ AGNs appear to be heavily obscured \citep[e.g.][]{Matthee24_LRD, Scholtz23_oscuredAGN} and are unlikely to dominate the UV emission.

An alternative explanation is the presence of a stochasticity in the galaxies luminosities: the relation between the UV magnitude of a galaxy ($\muv$) and the mass of the dark matter halo in which it resides ($M_h$) may not be univocal, but rather characterized by a dispersion $\sigma_{UV}$ \citep[e.g.][]{Mason23, Mirocha23, Shen23, Kravtsov24}. In this way, the increased normalization of the LFs can be explained by the up-scattered luminosities in the numerous low-mass halos dominating the high-$z$ population.

A physical mechanism that can naturally produce a scatter in the UV luminosities of galaxies is a bursty star formation history (SFH).
The complex interplay between multiple mechanisms, such as internal feedback processes, the continuous inflow of gas from the cosmic web and mergers, can cause a stochastic behavior in the star formation rate (SFR) of high-$z$ galaxies, making their UV luminosity fluctuate over time \citep[e.g.][]{Sparre17,Furlanetto22,Sun23a}.
Both JWST observations and cosmological simulations suggest that star formation in the first billion years after the Big Bang may have occurred in a particularly bursty fashion.

From the observational side, deep JWST surveys are revealing a broad range of SFRs in high-$z$ star forming galaxies \citep[e.g.][]{Curti23b, Endsley23b}. In particular, \citet{Endsley23b} recently found that UV-faint galaxies ($\muv \gtrsim -17$) often have weaker nebular emission lines and can be experiencing a down-turn in star formation compared to UV-bright galaxies ($\muv \lesssim -20$) which are more likely to have strong emission lines and to be in an up-turn of star formation. This is consistent with a picture whereby the bright-end of the UV LF is dominated by galaxies which have been up-scattered, whereas fainter galaxies show a range of star formation history stages. 
Moreover, the first detections of quenched low-mass galaxies ($M_\star<10^9\msun$) at $z>5$ \citep{Looser23, Looser23b, Strait23} can be interpreted as post-starburst galaxies undergoing temporary periods of quiescence due to a highly bursty and feedback-regulated star formation \citep{Gelli23,Gelli24, Dome23}.
The timescales on which these SFR variations occur tightly depend on the ongoing feedback processes within the galaxy \citep[e.g.,][]{Iyer2020}. Thus, constraining the level of stochasticity and burstiness of high-$z$ systems is essential for understanding the physical mechanisms driving their evolution.

State-of-the-art cosmological simulations also predict a highly time variable star formation in the early Universe \citep[e.g.][]{Ma18, Pallottini2022}.
However, whether the stochasticity in the SFRs of early galaxies is sufficient to explain the high number densities observed is still debated: \cite{Pallottini23} argue that an additional source of luminosity is needed already at $z\sim8$, while \cite{Sun23b} finds that $z\sim 8-12$ LFs can be reproduced with \code{FIRE-2} simulations, which show strongly time-variable SFHs.
Meanwhile, analytical studies have shown that, to match the observed LFs, the level of scatter in the UV luminosities $\sigma_{UV}$ needs to increase significantly above $z\gsim10$ \citep{Munoz23, Shen23, Mason23, Kravtsov24}.
In these works, the UV scatter is assumed to be independent of the halo mass. However, low-mass halos are intrinsically characterized by shallower potential wells and the galaxies residing in them are expected to be more sensitive to all those feedback and environmental processes that drive bursty star formation \citep[e.g.][]{Gelli20,Furlanetto22,Legrand22,Hopkins23}. For this reason, we expect stochasticity to be more relevant for lower mass halos, which, in a $\Lambda$CDM Universe, dominate the galaxy population at progressively higher-$z$.

In this paper, we develop a simple analytical model to explore a \textit{mass-dependent} UV scatter, i.e. increasing towards lower halo masses, independent of the redshift, described in Section~\ref{sec:model}.
We analyze the impact of this mass-dependent UV scatter on the LF, but also on other key observables: the neutral fraction in the intergalactic medium (IGM), the galaxy bias and spectral features, providing unique independent tools to help us constrain the relevance of bursty star formation in high-$z$ galaxies. These results are presented in Section~\ref{sec:results}. We discuss our results in the context of the $z>12$ LFs in Section~\ref{sec:sfe_discussion} and present our conclusions in Section~\ref{sec:conc}. %Detailed paper structure??
We assume a flat $\Lambda$CDM  cosmology with $\Omega_m=0.3$, $\Omega_\Lambda=0.7$ and $h=0.7$.

\section{Modeling the mass-dependent UV scatter}
\label{sec:model}

\begin{figure}[t!]
    \includegraphics[width=0.47\textwidth]{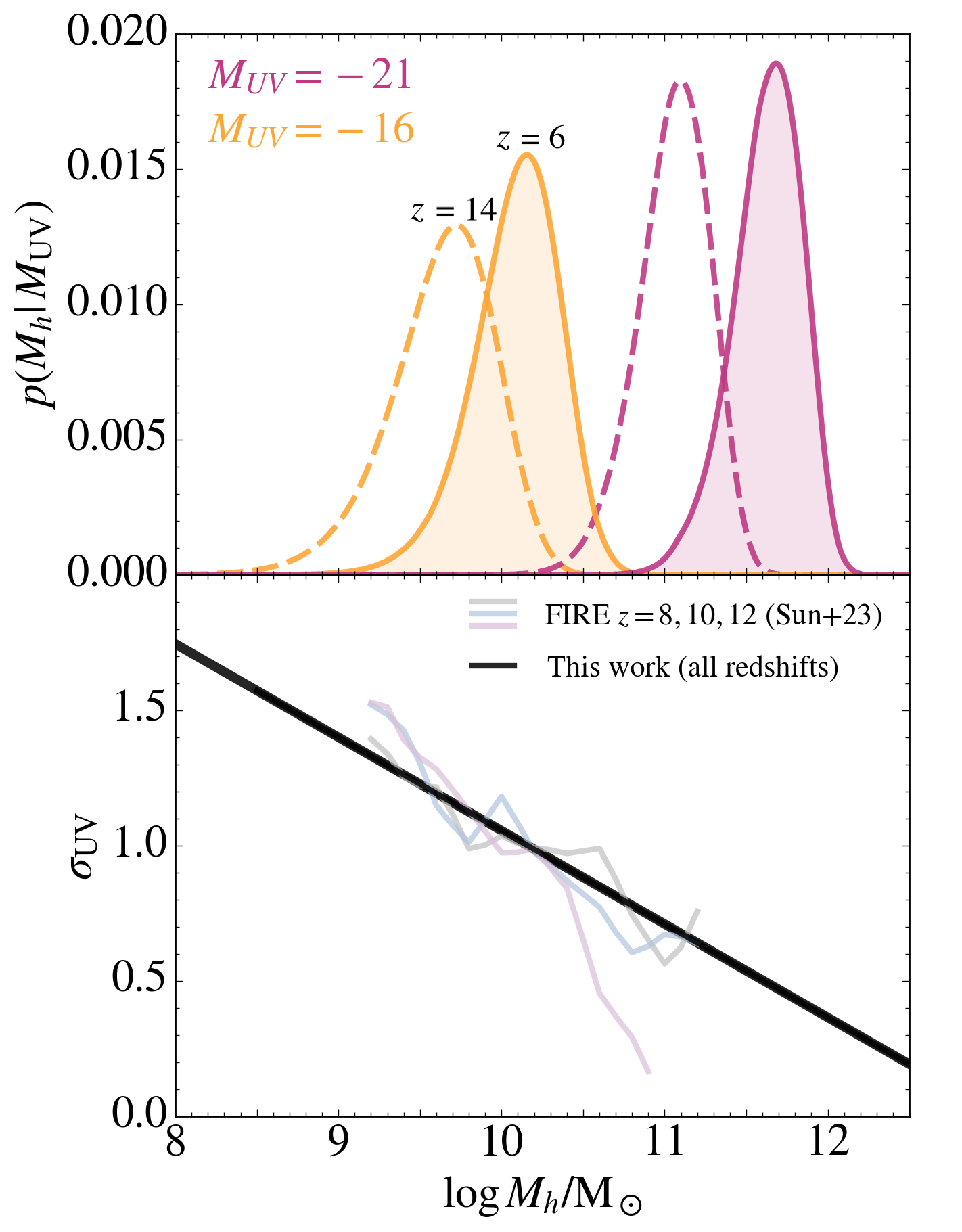}
    \caption{\textit{Top panel:} Probability distribution of halo masses at given UV magnitude $\muv$, with  corresponding scatter $\sigma$ determined by the model in the bottom panel. The distributions are shown for three different redshifts, highlighting the broadening of the distributions towards earlier epochs.
    \textit{Bottom panel:} Mass-dependent UV scatter $\siguv(M_h)$ as a function of the halo mass $M_h$. The black line shows the fiducial model used in this work, for all redshifts.
    \label{fig:sigmauv}
    }
\end{figure}

We model the stochasticity by assuming a mass-dependent but \textit{redshift-independent} scatter $\siguv(M_h)$, describing the dispersion in the $\muv-M_h$ relation, which decreases with halo mass.
This is motivated by cosmological simulations such as FIRE-2 which find a decreasing trend with the halo mass, without a significant evolution between $z\sim8-12$ \citep{Sun23b}.
We assume the scatter can expressed as a power-law relation $\siguv(M_h) = a\log M_h/\msun + b$.
We assume a fiducial model with $a=-0.34$ and $b=4.5$, shown in the bottom panel of Fig.~\ref{fig:sigmauv}.
We set these values after testing multiple sets of parameters $a$ and $b$ which best reproduce the shape of the UVLFs at $z\sim5-14$. This parameterisation of the scatter produces a similar trend to that seen in the FIRE-2 simulations results at $z\sim8-12$ (derived from Fig.~1 of \citealt{Sun23b}), which we show for comparison.
% We assume a fiducial model with $a=-0.34$ and $b=4.5$, shown in the bottom panel of Fig.~\ref{fig:sigmauv}, obtained by fitting the results of the FIRE-2 simulations at $z\sim8$ (derived from Fig.~1 of \citealt{Sun23b}). 
Interestingly, this scaling with halo mass is very similar to the $M_h^{-1/3}$ scaling expected if the stochasticity scales with the inverse of halos' escape velocities -- i.e. more tightly bound halos can retain more gas and have lower stochasticity.

\begin{figure*}[t!]
    \includegraphics[width=\textwidth]{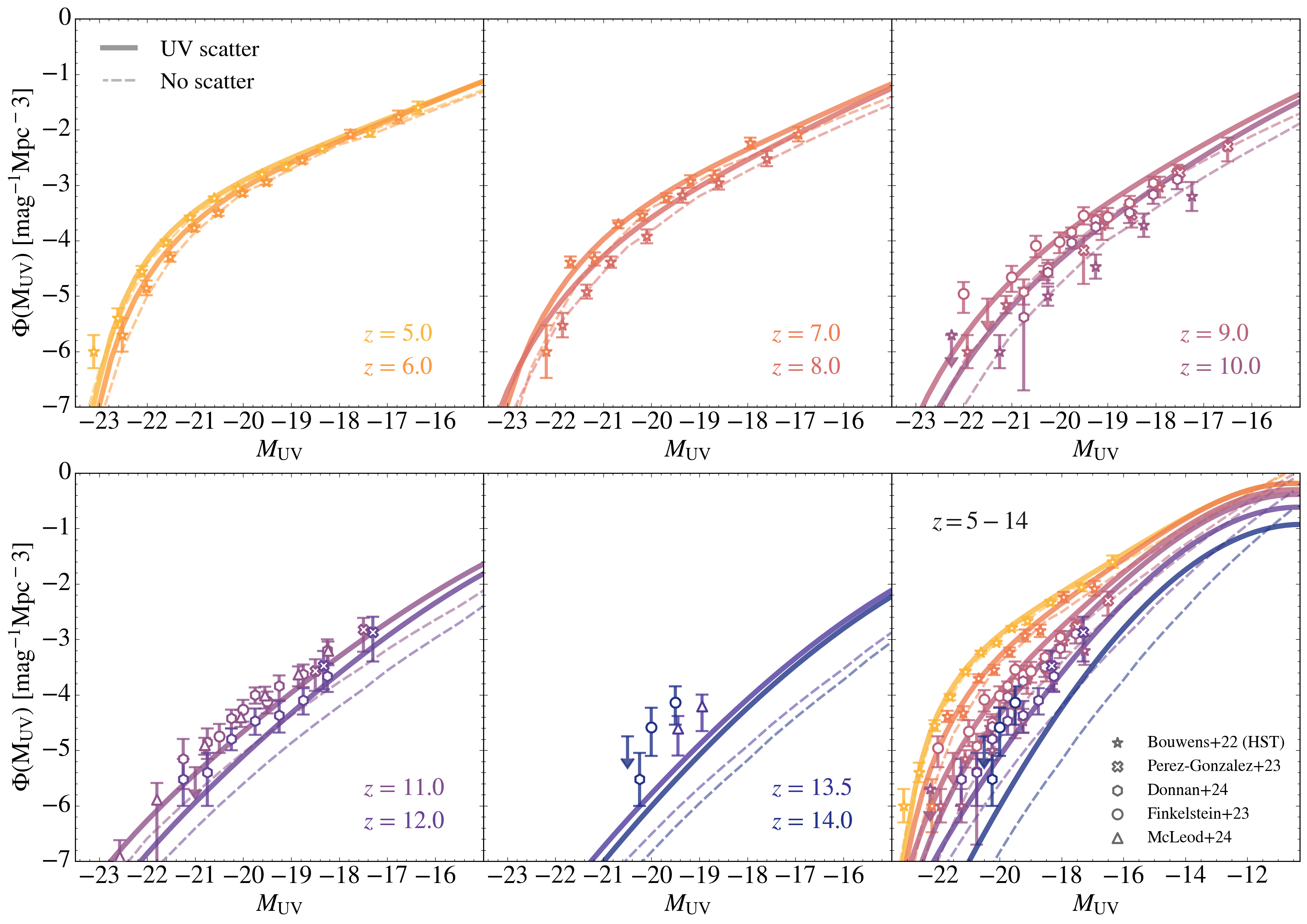}
    \caption{UV LF for the mass-dependent UV scatter model (solid lines) and the no-scatter model (dashed lines) at redshifts from $z=5$ to 14. Also shown are observations by \cite{Bouwens22lf,Finkelstein23, PerezGonzalez23,Donnan24, McLeod24}.
    \label{fig:lf}
    }
\end{figure*}
To include the scatter in our modeling, we adopt the conditional luminosity function (CLF) approach, following \cite{Ren19} \citep[see also][]{Whitler20}. The probability of a dark matter halo with mass $M_h$ to host a galaxy with magnitude $\muv$ is given by:

\begin{multline} 
    p(\muv \mid M_h)= \\ \frac{1}{\sqrt{2\pi}\siguv(M_h)} \exp \left(\frac{-[\muv - M_{{\rm UV}, c}(M_h, z)]^2}{2\siguv^2(M_h)}\right),
    \label{eqn:Muv_Mh_prob}
\end{multline}

where $M_{{\rm UV}, c}$ is the median magnitude at a given redshift $z$ and halo mass $M_h$.
We derive it by following \cite{Mason15,Mason23}, assuming 
$\langle {\rm SFR} \rangle \propto \epsilon_\star \, f_b M_h$, where $f_b$ is the cosmic baryon fraction and $\epsilon_\star $ the star formation efficiency, and then using \code{starburst99} \citep{starburst99}\footnote{We use \code{padova} stellar tracks \citep{Bertelli94} and assume a \cite{Kroupa01} initial mass function between 0.1 and 100$M_\odot$. However we note that the following results are not sensitive to these specific choices.}
to derive $\muv$.
We assume a mass-dependent star formation efficiency $\epsilon_\star(M_h) \sim  M_\star/M_h$, and we calibrate\footnote{We note that when introducing mass-independent scatter, \citet{Ren19} required a break in $\epsilon_\star(M_h)$ above $M_h \gtrsim 10^{11.5}$ to match the bright-end of the LF, however, this is not necessary in our model as $\siguv(M_h)$ is so low at these high halo masses.} 
it at a single redshift $z\sim5$ to reproduce the observed UVLFs by \cite{Bouwens23}
\citep[see also][]{Mason15, Ren19, Mirocha23}. Like the UV scatter, once calibrated, the derived star formation efficiency is assumed to be \textit{redshift independent}.

We can derive halo mass distributions $p(M_h \mid \muv)$ using a numerical approach, evaluating the probability in Equation \ref{eqn:Muv_Mh_prob} for an array of halo masses at fixed $\muv$ and $z$ \citep[see also][]{Whitler20}. The results are shown in the top panel of Fig.~\ref{fig:sigmauv} for two different magnitudes and redshifts.
The position of the peak of the distribution is determined by the median $M_{{\rm UV}, c}(M_h, z)$, while the width of the distribution is determined by the scatter $\siguv(M_h)$ and is independent of the redshift.
At fixed redshift, brighter galaxies are more likely to be hosted by larger dark matter halos. Note however that here we are not weighting by the halo mass function. As lower mass halos are in reality more numerous, galaxies we observe at fixed UV magnitudes will be more likely to be hosted in lower mass halos, as it will be evident in the next section.
The dispersion around the median $M_h$ is larger for lower luminosities. This is a direct consequence of the increasing $\siguv$ towards lower masses, making lower mass halos have high probability to appear with a wider range of luminosities.

Comparing the different redshifts, we can see that there is a broadening of the distribution towards high-$z$, meaning that galaxies at a given $\muv$ can be hosted at higher $z$ by a broader range of halos with respect to their lower $z$ analogues.
This behavior is due to the combined effect of: (i) the increasing median luminosity $M_{{\rm UV}, c}(M_h, z)$ with the redshift -- i.e. at fixed halo mass, galaxies are on-average more luminous at higher redshift, due to higher SFRs, and (ii) the increasing scatter $\siguv$ in lower mass halos. 

In the following sections we determine and analyze the expected impact of such mass-dependent UV scatter on different observables (UVLFs, SFR density, neutral hydrogen fractions, spectral features).

\section{Results} \label{sec:results}

We present our model\footnote{Tabulated version of the UVLF and luminosity density predictions are available online at: \url{https://github.com/violagelli/Stochastic_model_UVLF}} for the UV luminosity function in Section~\ref{sec:lf} and the luminosity density in Section~\ref{sec:lum_density}. In Section~\ref{sec:pred} we make predictions for other independent observables: galaxy clustering, the reionization timeline, and galaxy SEDs; which could be combined with the LF to constrain the level of stochasticity. 

\subsection{Luminosity function} \label{sec:lf} 
Following the CLF approach, we derive the luminosity functions by integrating the number density of dark matter halos, weighted by the probability that a halo $M_h$ hosts a galaxy with magnitude $\muv$:
\begin{equation}
    \Phi(\muv) = \int_{M_{h,min}}^\infty d M_h \, p(\muv \mid M_h) \, \frac{dn}{d M_h} ;
    \label{eqn:lf_clf}
\end{equation}
where $p(\muv \mid M_h)$ is given by Equation \ref{eqn:Muv_Mh_prob}, and for $dn/d M_h$ we use the halo mass function (HMF) from \cite{Reed07}.
We set a lower limit to the integral, $M_{h,min}$, to prevent the unphysical effect of very low-mass halos forming stars with extremely large UV scatter.
Its value is set at each redshift as the minimum halo mass for which atomic cooling can occur, i.e. corresponding to a virial temperature of $T_{\rm vir}\sim10^4\rm K$ \citep[e.g.][]{BrommYoshida11}. Following \citet{Mason15,Mason23} we add dust attenuation using the \citet{Meurer1999} attenuation law $A_\mathrm{UV} = 4.43 + 1.99\beta$, including the UV slope $\beta(\muv, z)$ empirically from observations by \citet{Bouwens2014}. We remove dust attenuation at $z>10$ motivated by recent results suggesting dust attenuation is negligible in $z>10$ galaxies \citep{Topping2024,Cullen2024,Morales2024}, though we note that beyond $z\gtrsim10$ dust has only a small impact on the UV LFs \citep{Mason23}.

The resulting LFs for our mass-dependent UV scatter model are shown in Fig.~\ref{fig:lf} from $z=5$ to 14.
We compare them with the model without UV scatter (dashed lines), where the evolution of the LF directly follows the evolution of the underlying halo mass function. The solid curves are the LF obtained including the mass-dependent scatter $\siguv(M_h)$ described in the previous section.

At progressively higher redshifts, the scatter in the UV luminosities has a stronger impact on the LF, which deviates more and more from the no-scatter case.
The presence of a UV scatter that decreases with halo mass \textit{naturally produces} larger number densities of $\muv \lesssim -14$ galaxies at higher redshifts with respect to the no-scatter case. This is due to low mass halos (characterized by higher UV scatter) being progressively more dominant at earlier epochs.

As evident from the bottom right panel of Fig.~\ref{fig:lf} where we show the LFs up to fainter luminosities, the model also predicts a faint end turnover of the LF at magnitudes $\muv\gsim-13$. This is because we have imposed a cut-off in galaxy formation at the atomic cooling limit $M \sim 10^8\,M_\odot$, i.e. lower mass halos cannot host a galaxy and thus will not be scattered up to populate the LF. While galaxies in higher mass halos are scattered down to lower luminosities during lulls in star formation, their number densities are lower than low mass halos, and they cannot boost the faint-end of the LF. If ever probed with deep observations \citep[e.g. in lensed fields with JWST][]{GLIMPSE}, this could provide an additional constraint on the dominant driver of star formation at high redshifts.

We find the mass-dependent UV scatter model successfully reproduces the observed HST and JWST up to $z\sim12$ \citep{bouwens:2022,Finkelstein23,PerezGonzalez23,Donnan24,McLeod24}, producing a $\sim1$\,dex increase in number densities of $\muv < -18$ galaxies at $z\gtrsim9$ compared to the no-scatter model. The model slightly overpredicts the bright-end of the LF at $z\sim8$, which is likely due to a combination of the dust modelling and the precise form of $\sigma_{\rm UV}(M_h)$.
However, at redshifts greater than $>12$, even the additional contribution from an enhanced UV scatter at low masses is not sufficient to reproduce the observed high number densities of galaxies. 
This suggests that additional physical processes may be at play at these redshifts to enhance the UV luminosities, as we will discuss in Sec.~\ref{sec:sfe_discussion}.

\begin{figure}
    \includegraphics[width=0.49\textwidth]{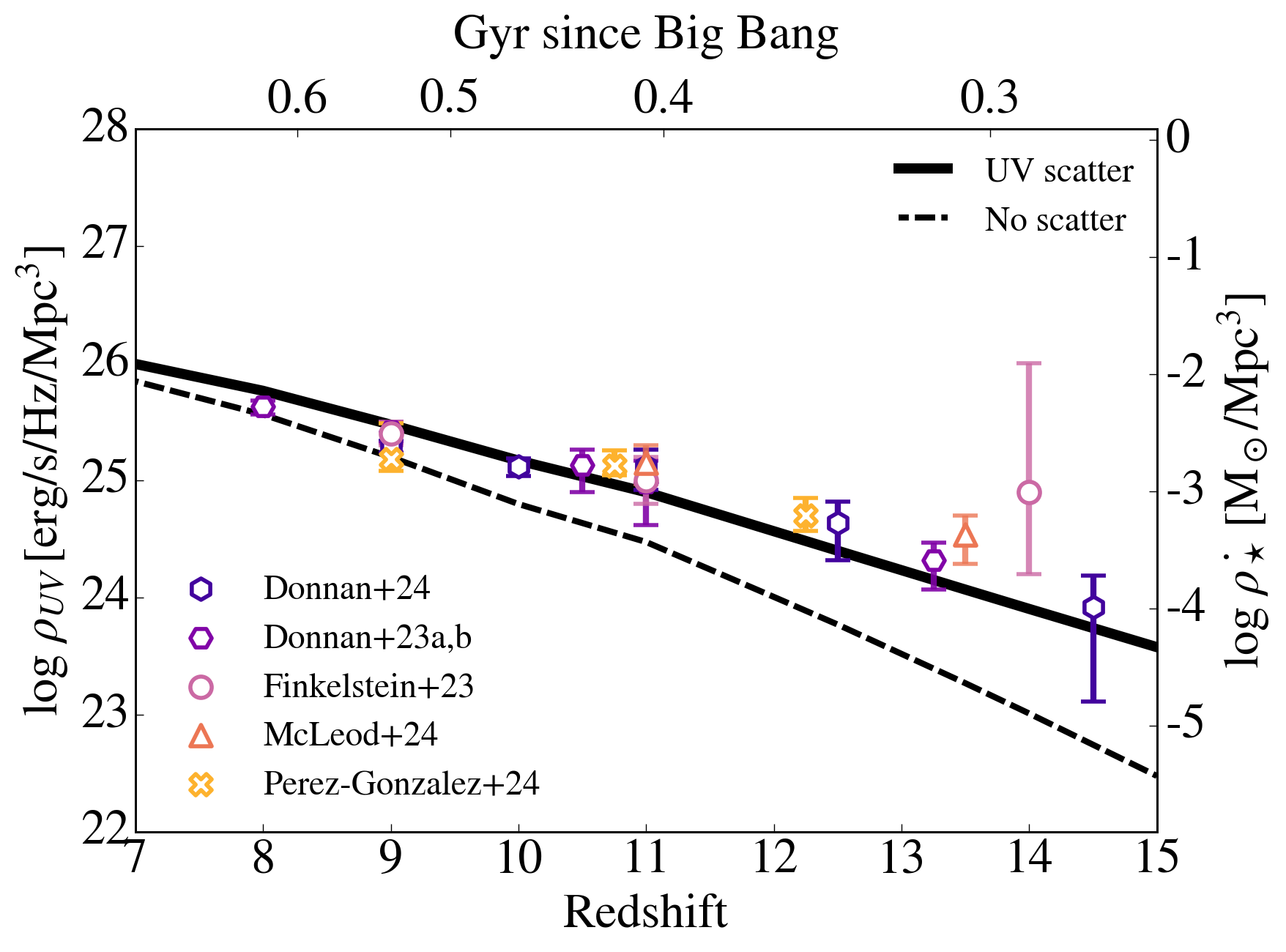}
    \caption{Luminosity density $\rho_{\rm UV}$ and cosmic star formation rate density $\rm SFRD$ as a function of redshift, derived integrating the UV LFs down to $M_{\rm lim}=-17$. The solid and dashed curves show the mass-dependent UV scatter and no-scatter models respectively. Also shown are observations from \cite{Bouwens22lf, Finkelstein23, PerezGonzalez23, Donnan24, McLeod24}. We note that the apparent match to the observations at $z\sim14$ is due to the steep faint-end slope of our LFs relative to that assumed by \citet{Donnan24} (see Section~\ref{sec:lum_density}).
    \label{fig:sfrd}
    }
\end{figure}

\subsection{Luminosity density evolution}\label{sec:lum_density}

By integrating the LFs we derive the luminosity density $\rho_{\rm UV}$ for the mass-dependent UV scatter model, shown in Fig.~\ref{fig:sfrd} as a function of redshift.
To be consistent with observations, we obtain $\rho_{\rm UV}$ by integrating the LFs down to a magnitude limit of $M_{\rm lim}=-17$. We also calculate the corresponding cosmic star formation rate density (SFRD) using the empirical relation $ \rm SFR / (\msunyr)= 8\times 10^{27} L / (erg\, s^{-1} Hz^{-1})$ from \cite{Madau99}.

The mass-dependent UV scatter and the no scatter models, shown with solid and dashed lines respectively, predict similar luminosity density at $z\sim7$, but progressively diverge towards higher redshifts.
While the no-scatter model underpredicts the observed luminosity density already from $z>9$, the mass-dependent UV scatter model is in good agreement with the latest JWST observations up to $z\sim11$, but then diverges from the observations at progressively earlier epochs.

We note that the luminosity density of the mass-dependent UV scatter model is consistent with the data from \cite{Donnan24} up to $z\sim14$, despite their LF exhibiting higher values at this redshift.
This happens because their double power-law fit yields a shallower LF slope than our model towards fainter magnitudes, resulting in lower luminosity density when integrating up to $M_{\rm lim}=-17$. It is important to note that the luminosity density only reflects the integrated LF and caution should be taken when drawing conclusions based solely on its value.
Therefore, while our luminosity density only slightly diverges from the data towards higher-$z$, this still suggests the need for extra luminosity production beyond that provided by stochastic star formation.

\subsection{Constraining the level of stochasticity} \label{sec:pred}
Even if it cannot fully explain the number density of $z\gtrsim12$ galaxies, bursty star formation still plays a role at cosmic dawn and it results from the complex interplay of multiple feedback processes acting on different timescales, which are not yet fully understood. Constraining the level of stochasticity in galaxies would offer key insights into the physical and feedback processes driving the evolution of galaxies of different masses across cosmic time, and their associated timescales \citep[e.g.,][]{Iyer2020}.

However, the LF and luminosity density are intrinsically degenerate with respect to multiple physical effects, such as an enhancement of the star formation efficiency or of the UV scatter. As discussed by e.g., \citet{Mirocha20,Munoz23}, this makes it impossible to use them alone as a tool to disentangle these physical processes.
For this reason, it is important to identify additional key observables that can help us to independently constrain the importance of stochastic star formation at high-$z$.

To this aim, we discuss the expected implications of mass-dependent UV scatter on multiple quantities: galaxy clustering (Section~\ref{sec:clustering}), the evolution of the intergalactic medium (IGM) neutral fraction (Section~\ref{sec:reionization}), and spectral energy distributions (Section~\ref{sec:SEDs}), providing predictions and tools to test the model against future observations.

\subsubsection{Clustering} \label{sec:clustering}
\begin{figure}[t!]
    \centering
    \includegraphics[width=0.48\textwidth]{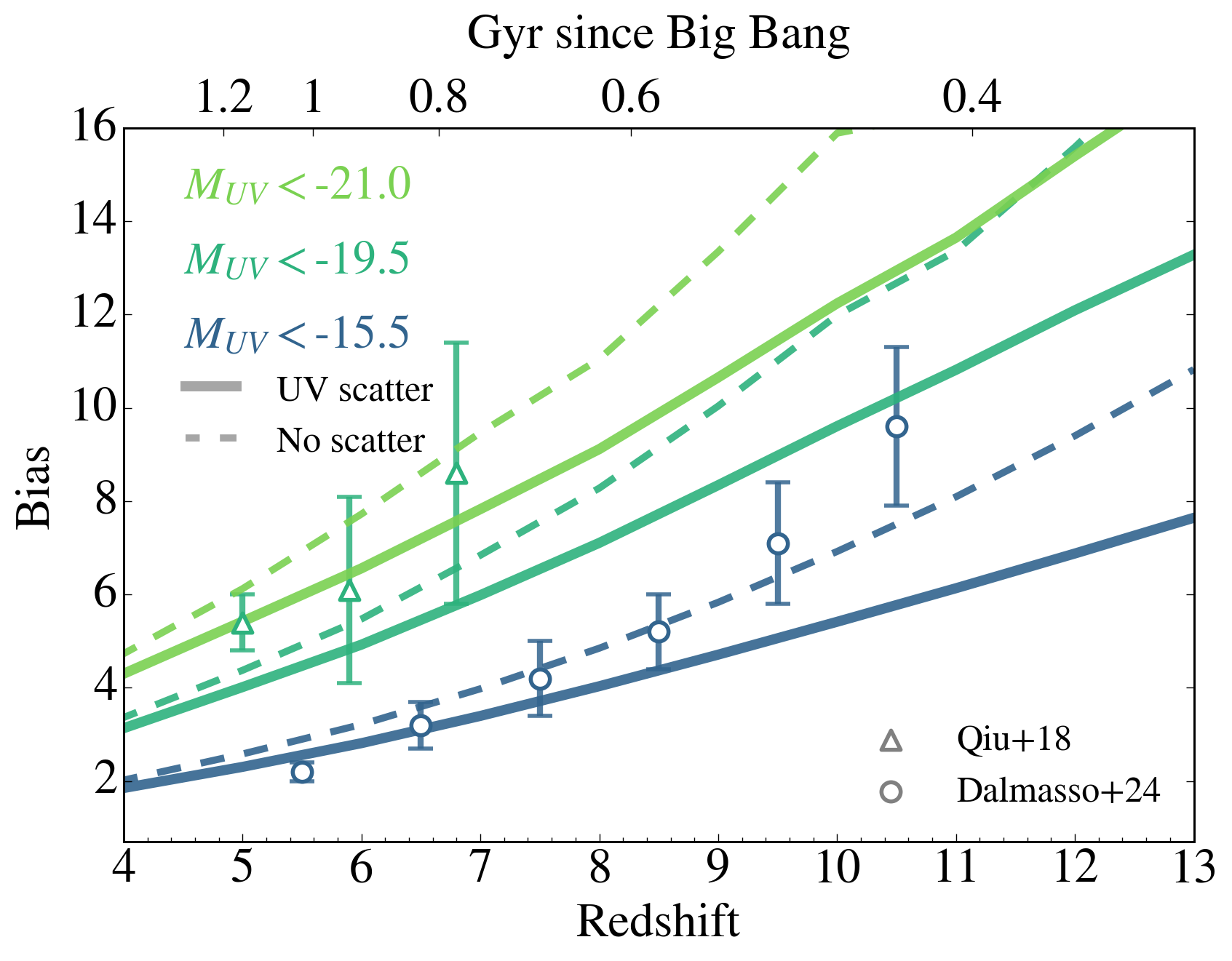}
    \caption{Redshift evolution of the predicted galaxy bias for the mass-dependent UV scatter model (solid lines) and the no-scatter model (dashed lines). Observations by \cite{Qiu18} and \cite{Dalmasso24} are shown.
    \label{fig:bias}
    }
\end{figure}
A key observational test of the importance of stochasticity in high-$z$ galaxies is to connect galaxies to their dark matter halos.
The clustering of galaxies is driven by the underlying distribution of the dark matter halos that they inhabit \citep[see e.g.,][for a recent review]{Wechsler2018}.
As discussed by \citet{Munoz23}, constraints on galaxy clustering can thus break the degeneracies of the UVLF and distinguish the presence of UV stochasticity from the effect of an increased average UV emission.

In a $\Lambda$CDM Universe, the galaxy bias (quantifying the halo-galaxy connection) is expected to increase with the halo mass: galaxies hosted by massive halos cluster more strongly than those residing in lower mass ones. Estimates of Lyman-break galaxy clustering at $z\sim4-7$ have revealed the brightest galaxies tend to reside in the most massive halos \citep[e.g.,][]{Barone-Nugent14,Harikane2016,Harikane2022,Qiu18,Dalmasso2024a}. 
As discussed by \citet{Munoz23}, scatter in the $\muv-M_h$ relation will lower the bias of galaxies, since UV luminous galaxies can be found in a broad range of halo masses.

It is therefore important to provide predictions for the galaxy bias in our mass-dependent UV-scatter model.
We estimate the luminosity-weighted effective bias of galaxies as \citep[see e.g.,][]{Munoz23}:
\begin{equation}
    b_{\rm eff}(\muv) = \frac{1}{\Phi(\muv)} \int d M_h \, p(\muv \mid M_h) \, \frac{dn}{d M_h} b(M_h)
\end{equation}

where we use the linear bias $b(M_h)$ fits from \cite{Tinker10}.
Our predictions for the redshift evolution of the bias at fixed $\muv$ are shown in Fig.~\ref{fig:bias}.

As expected, biases from the UV scatter model always have lower values than the no-scatter case, and the difference is progressively more pronounced towards higher redshifts as galaxies of fixed $\muv$ occupy lower mass halos. 
As discussed by \citet{Dalmasso2024a}, measurements of galaxy bias are extremely challenging at high-$z$ as estimates are limited by both Poisson noise due to low number counts and cosmic variance in small fields. 
Pre-JWST studies have measured the bias up to $z\sim7$ for $\muv \lesssim -19$ galaxies, but the errors are too large to draw definite conclusions and constrain the models, as we can see when comparing our model with data from \cite{Qiu18}.

However, the sensitivity of JWST makes more accurate estimates of the bias feasible by enabling us to measure clustering for sub-$L^\star$ galaxies at $z>6$ for the first time \citep[e.g.,][]{Endsley2020,Dalmasso24}. 
We compare our results with those of recent estimates from JWST/NIRCam observations 
of $\muv \lesssim -15$ galaxies at $z\sim5-10$ in the GOODS-S field by \cite{Dalmasso24}.
The bias estimated by \citet{Dalmasso24} shows a very steep increase with redshift. At $z<9$ the difference between our two models is too small to be probed by these observations. However, $z>9$, the \citet{Dalmasso24} constraints are more consistent with the no-scatter case, and even slightly above both models at $z\sim10$.

This could suggest galaxies experience a lower level of stochasticity than what we have assumed and/or are hosted in higher mass halos than expected at higher redshifts. This is somewhat surprising, as at fixed UV magnitude, models predict and clustering results imply that galaxies reside in lower mass halos at higher redshifts as star formation rates are higher \citep[e.g.,][]{Mason15,tacchella:2018,Harikane2022}. This may potentially imply a change in the star formation efficiency with halo mass at high redshift. 

However, we note that \citet{Dalmasso24} assumed a different cosmology than we have, and full interpretation of JWST observations likely requires a Halo Occupation Distribution (HOD) model approach \citep[e.g.,][]{Harikane2022} to separate non-linear clustering on small scales due to satellite galaxies, as noted by \citet{Dalmasso24}. In addition, large area, unbiased, surveys with JWST will be essential to constrain models by probing the number density and clustering of UV bright galaxies ($\muv \lesssim -20$) at $z\gtrsim7$. 

\begin{figure*}[t!]
    \centering
    \includegraphics[width=\textwidth]{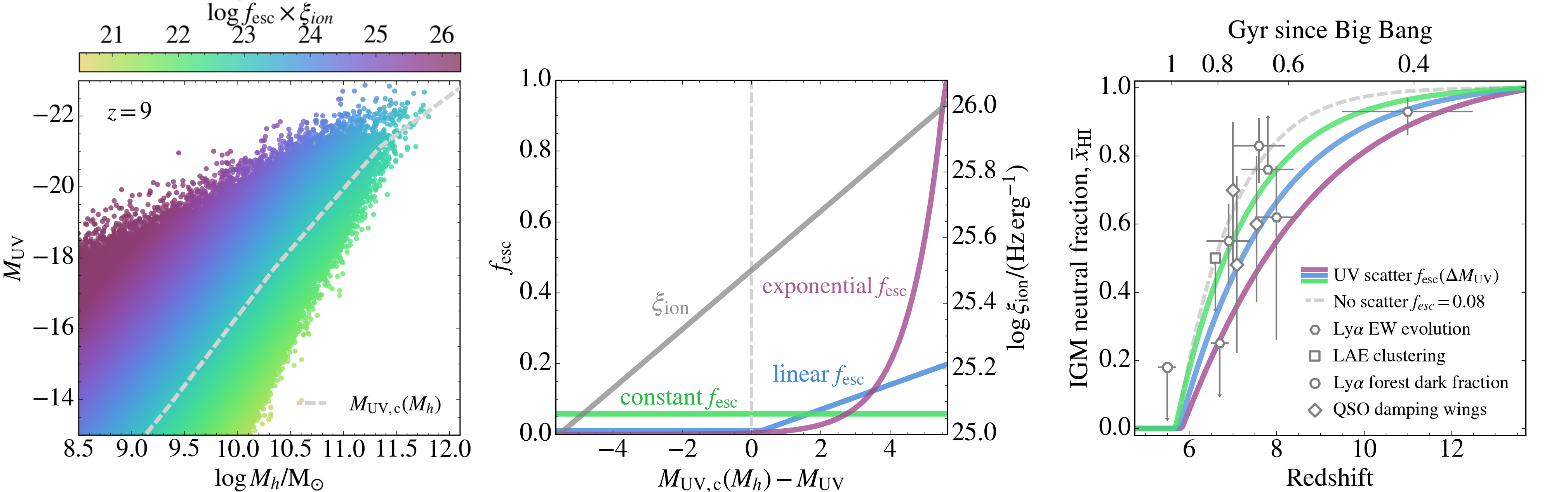}
    \caption{
    {\it Left}: $\muv-M_h$ relation at $z=9$, populated with galaxies following the halo mass function and coloured with the product $f_{\rm esc}\xi_{\rm ion}$ (assuming an exponentially increasing $f_{\rm esc}$ and linearly increasing $\xi_{\rm ion}$ with $\Delta \muv$), showing how up-scattered galaxies will contribute the most to the number density of ionizing photons.
    {\it Centre}: escape fraction $f_{\rm esc}$ and ionising photon production efficiency $\xi_{\rm ion}$ as a function of $\Delta \muv$. For the escape fraction we show three different cases of constant, linear and exponential increase towards up-scattered galaxies.
    {\it Right}: Redshift evolution of the fraction of neutral hydrogen in the IGM for the same three scenarios, as well as for the case without scatter, assuming fixed $f_{\rm esc}=0.08$ and $\log_{10}\xi_\mathrm{ion}/ (\rm Hz\, erg^{-1})=25.7$. We also plot measurements from observations of: the evolution of the $\rm Ly\alpha$ equivalent width (EW, \citealt{Mason18a, Mason19a, Bolan22, Nakane23}), the clustering of $\rm Ly\alpha$ emitters (LAE, \citealt{Ouchi10, Sobacchi15}), $\rm Ly\alpha$ forest dark pixel fraction \citep{McGreer15}, and quasar damping wings \citep{Davies18, Greig19}.   
    \label{fig:reionization}
    }
\end{figure*}

\subsubsection{Reionization history} \label{sec:reionization}

Constraints on the IGM reionization process offer vital information about star formation in \textit{all} sources, even those too faint to detect directly with JWST.
\citet{Munoz24} recently described how early JWST estimates of high ionizing photon production efficiency in high-$z$ galaxies, taken at face-value, may imply a potential `overproduction' of ionizing photons, which is at odds with current constraints on the timeline of reionization. Bursty star formation may alleviate this tension.
In particular, \citet{Munoz24} assumed constant values of the ionizing photon escape fraction $f_{\rm esc}$ and ionizing photon production rate, $\xi_{\rm ion}$ at fixed $\muv$, which may not be realistic assumptions in a stochastic star formation scenario (see also Section~\ref{sec:SEDs}).

Here, first, we discuss results from observations and simulations implying that ionizing photon production and escape may be linked to bursty star formation. Then, we combine our mass-dependent UV scatter model with simple models for $f_{\rm esc}$ and $\xi_{\rm ion}$ as a function to offset from the median $M_\mathrm{UV,c}(M_h)$ relation to explore how bursty star formation impacts the reionization history.

Observationally, high ionizing photon production rate ($\xi_{\rm ion}$) is linked to high equivalent width nebular emission lines, and thus young, massive stars \citep[e.g.,][]{Chevallard2018,Tang2019,Tang2021}. Therefore, if galaxies undergo bursty star formation, the number of ionizing photons they produce will vary significantly throughout their star formation history. Indeed, as discussed in the introduction, \citet{Endsley23b} recently found that UV-faint galaxies ($\muv \gtrsim -17$) are more likely to have weaker nebular emission lines and be experiencing a down-turn in star formation compared to UV-bright galaxies ($\muv \lesssim -20$). \citet{Endsley23b} discuss that one interpretation of the weaker emission lines in UV-faint galaxies is due to their star formation histories (rather than an increase in $f_\mathrm{esc}$ in faint galaxies, which produces a degenerate effect in photometric data). In this case, they find $\xi_{\rm ion}$ is a broad distribution that can be strongly dependent on UV magnitude.

Furthermore, hydrodynamical simulations predict that the escape fraction of ionizing photons may also coincide with (or shortly after) star formation bursts \citep{Kimm2014,Paardekooper2015,Ma2020a,Barrow2020}. In particular, hard radiation from massive stars may allow ionizing photons to escape by creating `porous' density-bounded HII regions and/or stellar winds or supernovae feedback can open low-density channels \citep[e.g.,][]{Zackrisson2013,Steidel2018}. 
From the observational side, this picture is not completely clear:
while many Lyman-continuum leakers detected at $z\sim0.3-0.4$ are in a period of intense star formation, 
with specific star formation rates $\gtrsim 10$\,Gyr$^{-1}$ and star formation rate surface densities $\Sigma_\mathrm{SFR} \gtrsim 10\,M_\odot$\,yr$^{-1}$\,kpc$^{-2}$ \citep[e.g.,][]{Izotov2016,Izotov2018,Flury2022}, 
observations at $z\sim3$ show no clear trend of $f_\mathrm{esc}$ with specific star formation rate, age or $\Sigma_\mathrm{SFR}$ \citep{Pahl2022,Pahl2023}. 

In the context of our mass-dependent UV scatter model, we combine these observational and theoretical insights with our UV LFs to assess the impact of stochasticity on the timeline of reionization.
In particular, as discussed by \citet{Tang2019}, \citet{Naidu2022} and \citet{Endsley23b}, among others, we consider that galaxies experiencing a burst of star formation, i.e. that will be up-scattered in the $\muv-M_h$ distribution, could have an outsized contribution to the ionizing photon budget.

This effect can be translated into $\xi_\mathrm{ion}$ and $f_{\rm esc}$ being described by a distribution at fixed UV magnitudes rather than fixed values for every galaxy.
The production rate of ionizing photons available for reionization can therefore be expressed as:
\begin{multline}
    \dot{n}_{\rm ion}(z) =  \int_{-\infty}^{M_{lim}} d\muv L(\muv)\Phi(\muv) \\
    \times 
    f_{\rm esc} \, \xi_{\rm ion} \, p(f_{\rm esc},  \xi_{\rm ion} \mid \muv) ;
\end{multline}
which can be solved when an empirical relation between $\xi_{\rm ion}$ and $\muv$ is available.

Within our modeling framework, we model $\xi_\mathrm{ion}$ and $f_{\rm esc}$ as functions of $\Delta\muv = M_{\rm UV,c}(M_h) - \muv$. 
As illustrated in the middle panel of Fig.~\ref{fig:reionization}, we assume $\log_{10} \xi_{\rm ion}$ increases linearly with $\Delta\muv$, which naturally produces a log-normal distribution of $p(\xi_{\rm ion} \mid \muv)$, similar to observations \citep[e.g.,][]{Prieto-Lyon23,Endsley23b}.
For the escape fraction we explore three different scenarios: (i) exponentially increasing $f_{\rm esc}(\Delta\muv)$, which produces an exponential distribution of $p(f_{\rm esc} \mid \muv)$ \citep[similar to that derived for $z\sim3.5$ star forming galaxies by][]{Kreilgaard2024}; (ii) linearly increasing $f_{\rm esc}(\Delta\muv)$; and (iii) constant $f_{\rm esc}(\Delta\muv)$ for all galaxies.

In Fig.~\ref{fig:reionization}, left, we show the $\muv-M_h$ relation at $z=9$, populated following the halo mass function and our mass-dependent $\sigma_\mathrm{UV}(M_h)$ model. Each galaxy is coloured with the corresponding product of escape fraction and ionizing efficiency for the exponentially increasing $f_{\rm esc}$ case.

Using these assumptions, we can derive $\dot{n}_{\rm ion}(z)$ as:
\begin{multline}
    \dot{n}_{\rm ion}(z) =  \int_{-\infty}^{M_{lim}} d\muv L(\muv)\Phi(\muv) \\
    \times 
    \int_{M_{h,min}}^{\infty} dM_h  \, f_{\rm esc}[\Delta \muv (M_h)] \, \xi_{\rm ion}[(\Delta \muv (M_h)]\\
    \times p(\muv \mid M_h) \, \frac{dn}{dM_h} ;
\end{multline}

where we set $M_\mathrm{lim}=-13$, but note that our results are relatively insensitive to the exact value of this choice due to the turn-over of the UV scatter LFs at $\muv \sim -13$ (Fig.~\ref{fig:lf}). 
The ionizing photon rate production can be then used to calculate the cosmic evolution of the ionized hydrogen fraction by solving Eq. (6) in \cite{Mason15}.

The resulting reionization histories for the three scenarios are shown on the right of Fig.~\ref{fig:reionization}. We also compare to the no-scatter case, for which we assume fixed $f_{\rm esc}$ and $\xi_{\rm ion} = 10^{25.7} \rm Hz\, erg^{-1}$ (consistent with the latest JWST results; e.g. \citealt{Endsley23,Atek24}).
For each model, $f_{\rm esc}$ is tuned to obtain the end of reionization by $z\simlt6$, as inferred from Lyman-alpha forest observations \citep{Qin2021,Bosman2022,Jin2023}. In the no-scatter scenario this leads to a value of $f_{\rm esc}=0.08$. As discussed by \citet{Munoz24} this may be in tension with estimates from low-$z$ analogs which would predict higher $f_{\rm esc}$.

When adding stochastic star formation, we obtain an earlier onset and more gradual evolution of reionization, as low mass halos can start producing significant ionizing photons at early times \citep[see also][]{Furlanetto22}. We find the shape of the reionization timeline is strongly dependent on our $f_\mathrm{esc}$ model.
The steeper the dependency of $f_\mathrm{esc}$ with $\Delta \muv (M_h)$ (i.e. going from the constant to the exponential scenario), the earlier the start and the shallower the evolution of the reionization process.

At face-value, stochastic star formation exacerbates the potential ionizing photon crisis: 
due to the higher luminosity densities in our scatter model, we require an even lower value of $f_{\rm esc}\sim0.05$ than the no-scatter case ($f_{\rm esc}\sim0.08$) to obtain a $z\sim6$ end of reionization. 
We note here that our linear $\xi_\mathrm{ion}$ model also assumes $\xi_\mathrm{ion}$ is higher for starbursting galaxies, but even when using a constant $\xi_\mathrm{ion}$ model the higher luminosity density of the scatter model implies a higher injection rate of ionizing photons at $z\gtrsim8$ compared to the no-scatter case.
This motivates the need to consider a distribution of escape fractions and $\xi_\mathrm{ion}$, different for galaxies in different stages of their bursty SFH, to properly model early galaxies' contribution to reionization.

We note however that while there are observational constraints on $\xi_\mathrm{ion}$ at high-$z$, the escape fraction is not directly constrained. Thus our predictions here are merely illustrative, but demonstrate that stochasticity can have a marked impact on the timing of reionization and is an important additional constraint to understand early star formation, and can be constrained now to $z\simgt10$ with JWST observations \citep[e.g.,][]{Nakane23,Umeda2023}. Indeed, recent analyses of CMB data have found higher values of the Thomson scattering optical depth compared to \citet{Planck18}, implying a possible early tail to reionization \citep{Pagano2020,deBelsunce2021,Giare2024}. As recently discussed by \citet{Asthana2024} an early start to reionization is not inconsistent with the requirement from the Lyman-$\alpha$ forest that reionization is complete by $z\sim5.3-6$ \citep{Bosman2022,Qin2021}.

As stochasticity implies lower mass halos (which are less biased, see above) are more important for reionization this would also impact the morphology of reionization. We would expect more numerous, smaller bubbles at fixed neutral fraction compared to a case with no stochasticity \citep[e.g.][]{Hutter21, Hutter2023, Lu24}. This could potentially be measured via Lyman-alpha transmission and 21-cm intensity mapping. 

\subsubsection{Spectral Energy Distributions} \label{sec:SEDs}

\begin{figure*}[t!]
    \centering
    \includegraphics[width=0.95\textwidth]{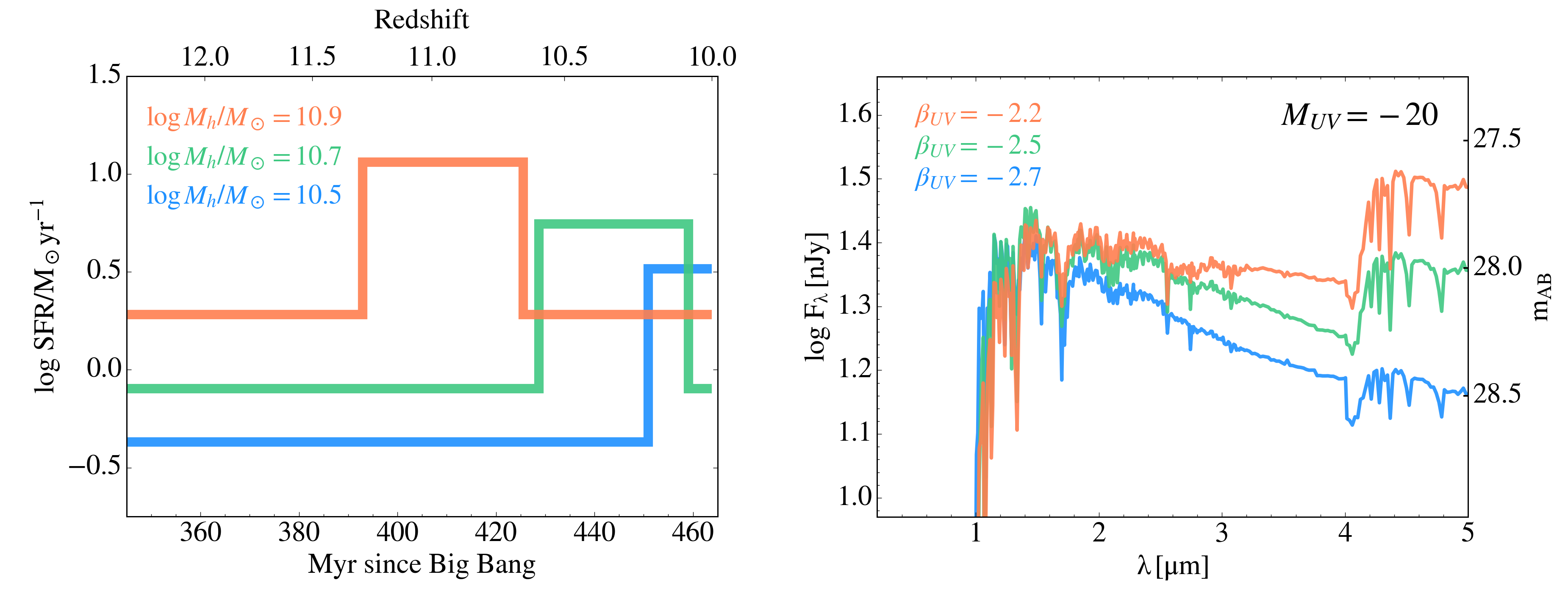}
    \caption{Examples of star formation histories (\textit{left}) and continuum spectral energy distributions (\textit{right}) for three galaxies at $z=10$ with same UV magnitude $\muv=-20.$
    \label{fig:sfh_sed}
    }
\end{figure*}

Highly time variable, bursty star formation histories that can produce drastic variability of the luminosity of a galaxy across its evolution, should also cause large scatter in galaxies' spectral features when we observe galaxies at fixed $\muv$ or stellar mass, as we are observing galaxies at different phases in their SFHs \citep[e.g.][]{Iyer2024,Sun23a,Endsley23b}. 
Different spectral features are sensitive to different timescales and galaxy properties, and are commonly used to probe galaxy star formation histories \citep[e.g.][]{Weisz12,Sparre17,Ciesla23,Cole23, Topping2024, Langeroodi24}.
In the context of understanding the role of stochasticity at high-$z$ it is important to identify which are the best observable features to help us constrain stochasticity 

JWST makes it possible to measure variations across galaxy populations at $z\sim5-15$. Several recent works have quantified stochasticity at $z\gtrsim5$ via the scatter in the star-formation main sequence using JWST photometry \citep[SFR - stellar mass, e.g.,][]{Ciesla23,Cole23,Caputi2023}, finding evidence for a potential increase in scatter with increasing redshift. However, it becomes increasingly challenging to measure accurate stellar masses from photometry for bursty galaxies, due to the `outshining' of rest-optical stellar emission by nebular emission from younger stars \citep[e.g.,][]{Stark2013,Whitler2023,GArteaga2023,Narayanan2024}. Analyses at fixed UV magnitude may provide more reliable comparisons with models during this bursty epoch of galaxy formation \citep[e.g.,][]{Endsley23b}.

To this aim, we explore the impact of simple bursty star formation histories on galaxy SEDs at fixed $\muv$, to avoid the issues of stellar mass uncertainties.

Following \cite{Mason15}, we set the average star formation rate of a halo of mass $M_h$ as $\langle {\rm SFR} \rangle \propto \epsilon_\star(M_h)\, f_b M_h$.
On top of this constant evolution we artificially add a ``burstiness" by modeling the SFH as a periodic pattern in which burst-phases and lull-phases alternates. Specifically the SFR varies between  $\rm \log \langle SFR\rangle + \Delta \log SFR$, during bursts, and $\rm \log \langle SFR \rangle - \Delta \log SFR$, during lulls. The value of $\rm \Delta SFR$ and the duration of burst and lull phases is calibrated so that the stellar mass assembled over the total period remains the same as for the constant evolution.

To check how this burstiness affects the emission of galaxies, we produce synthetic continuum SEDs.
We sample the SFH as a series of single bursts every 1~Myr and use \code{starburst99} \citep{starburst99} with \code{padova} stellar tracks \citep{Bertelli94} and assuming a \cite{Kroupa01} initial mass function between 0.1 and 100~$M_\odot$  to derive the rest-frame spectra and $\muv$ \citep[see also][]{Gelli21}.

We then select galaxies at the same magnitude $\muv$, finding that they can show a wide range of different SFH and properties due to their stochastic behaviour.

In Fig.~\ref{fig:sfh_sed}, left panel, we show as an example the SFHs of three galaxies with same magnitude $\muv=-20$ at $z=10$, but hosted in halos of different mass.
The one in the lowest mass halos is experiencing a burst of star formation, and is therefore up-scattered in the $\muv-M_h$ relation, whereas the ones in higher mass halo are in a lull phase.

In the right panel of Fig.~\ref{fig:sfh_sed} we show with the same color coding their redshifted observed flux $z=10$.
By definition, their rest-frame 1500\angstrom~ luminosity is the same, but we can see that significant differences are present in their UV slopes $\beta_{UV}$ and Balmer breaks at $\lambda_{rest} \sim 3645\angstrom$. Galaxies experiencing a recent rise in the SFR have steeper UV slopes, since their emission is dominated by young massive stars. On the other hand, galaxies undergoing a constant or recently decreasing period of SFR have shallower UV slopes and much more prominent Balmer breaks.

While not modelled here, we also expect the strength of nebular emission lines that trace star formation on short timescales ($<10$~Myr), such as $H\alpha$, to be enhanced during bursty phases \citep[see e.g.,][]{Weisz12,Sparre17}.
As a consequence, the $\beta_{UV}$, Balmer breaks and line equivalent widths represent quantities that can give us precious direct insight on bursty star formation \citep[see also e.g.][]{Endsley23b, Trussler24, Langeroodi24, Topping2024}.
Larger samples of galaxies are expected to show wide distributions of these quantities at fixed magnitude.
In a forthcoming work, we will extend our predictions, analyzing in detail the expected distributions of $\beta_{UV}$ slopes and Balmer breaks, as well as of emission line equivalent widths and other spectral features, all essential to constrain the bursty behavior of high-$z$ galaxies and to infer the timescales and physical processes regulating them.

\section{discussion: The abundance of galaxies at \MakeLowercase{$z>12$}}\label{sec:sfe_discussion}
Despite the success of the mass-dependent UV scatter model in naturally producing higher UVLFs at progressively higher redshifts, it still cannot reproduce the observed high number densities of galaxies at $z>12$.

Our analysis was based on some fundamental assumptions regarding: (i) the cosmological parameters, (ii) the halo mass function, and (iii) the form used for the matter transfer function. We tested how these assumptions impact the predicted LFs in Appendix~\ref{sec:cosmoassumptions}, and find that their effect is mostly negligible (with some exception for the transfer function).

We also tested the mass dependency of $\sigma_{\rm UV}(M_h)$, varying the slope and intercept of the power-law relation, but we find that no redshift-independent power-law form of $\sigma_{\rm UV}(M_h)$ can reproduce the LFs from $z=5$ to $z=14$ simultaneously.
Thus, the persisting discrepancy at $z>12$ suggests that stochastic star formation alone is not sufficient to fill the gap and that additional physical processes may be at play. 

As mentioned in the introduction, another way to boost the abundance of high-$z$ galaxies is through an enhancement of the average $\muv-M_h$ relation. This could be achieved if galaxies are forming stars in a more efficient manner at high-$z$.
Our model currently employs a redshift independent star formation efficiency, a strong assumption that likely needs to be relaxed to match the $z>12$ observations.

The fact that our model reflects a stochasticity proportional to the inverse of the halo escape velocity offers an interesting insight. As $v_\mathrm{esc} \sim M_h^{1/3} (1+z)^{1/2}$ we should expect a \textit{decrease} in stochasticity at fixed halo mass at higher redshifts, as halos are more compact and thus better able to retain their gas. This implies higher redshift halos should host higher gas masses at fixed halo mass, which may provide fuel for more efficient star formation.

As proposed by \citet{Dekel23}, a possible cause for higher star formation efficiency at high redshift is an increase in the density of star forming regions. Cloud-scale simulations show that when a surface density threshold is reached ($\Sigma \gtrsim 10^3 \, M_\odot$\,pc$^{-2}$) stellar feedback may not be fast enough to counteract gravitational collapse and the star formation efficiency can be extremely enhanced from usual values of $\simlt10\%$ to $\sim100\%$ \citep[e.g.][]{Grudic18,Lancaster21, Menon23, Menon24}. Galaxy-scale hydrodynamic simulations show that to create such high density clouds likely requires a turbulent gas-rich environment where gas is compressed by feedback-driven winds and shocks from inflowing gas \citep[][]{Ma20}. \citet{Dekel23} and \citet{Li23} discuss that we may expect more galaxies at high redshift to satisfy these high star formation efficiency conditions as halos become increasingly compact and accretion rates increase.

Indeed, observations show galaxies appear to be increasingly compact at $z>4$ \citep[e.g.][]{Langeroodi23,Morishita2024}. Furthermore, extremely high electron densities ($n_e\simgt10^{4}$ cm$^{-3}$) have been detected in two UV-bright, compact $z\simgt6$ sources (GNz11 and RXCJ2248-ID) which show UV emission line signatures of massive stars \citep[][though c.f. \citealt{Maiolino23_gnz11} where the high electron density of GNz11 is attributed to an AGN broad-line region]{Senchyna23,Topping24}. \citet{Topping2024} suggest such high electron densities could result from the high density birth clouds described above, which can fuel efficient star formation. Larger samples of $z>6$ sources with deep, high-resolution rest-UV spectroscopy trace both massive star signatures and ionized gas densities, combined with morphological information, should enable a better understanding of the links between efficient star formation and high densities.

\section{Conclusions} \label{sec:conc}
We have explored the impact of a mass-dependent scatter of the $\muv-M_h$ relation on the UV luminosity functions of high-$z$ galaxies, and other observables, to understand the relevance of bursty star formation in the early Universe. We assumed a decreasing trend $\sigma_{\rm UV}(M_h)\sim M_h^{-1/3}$ with increasing $M_h$, reflecting the higher sensitivity of low-mass halos to stochastic star formation processes due to their lower halo escape velocities.
Our main conclusions are the following:
\begin{itemize}
    \item The mass-dependent UV scatter model naturally produces higher UV LFs at progressively higher redshifts with respect to the no-stochasticity case, due to low-mass halos being more dominant at earlier epochs;
    \item The predicted LF and luminosity density are in agreement with the latest JWST observations up to $z\sim12$, but start to diverge at progressively earlier epochs, suggesting the need for additional UV luminosity production on top of that provided by stochastic star formation. This may be linked to the increasing compactness, and thus density, of halos at fixed mass with redshift;
    \item The model predicts significantly lower galaxy bias values than the no-scatter case as a function of redshift, with the models diverging at $z\gtrsim8$ for $\muv \sim -21$ galaxies;
    \item Assuming that up-scattered galaxies are the dominant source of ionizing photons yields an earlier start to reionization and a more gradual reionization timeline than a model without scatter;
    \item Balmer breaks, nebular emission line strengths and UV-slopes are also sensitive to bursty star formation histories and can have significantly different values for galaxies with same $\muv$.
\end{itemize}

Our results show that bursty star formation leading to a mass-dependent stochasticity in the UV luminosity - halo mass relation is expected to have an impact on multiple observable quantities.
JWST observations are probing the high-$z$ UV luminosity functions with increasing depth and precision, but, as discussed by \citet{Munoz23}, the contribution of bursty star formation cannot be constrained by the LF alone due to degeneracies with other physical effects.
For this reason, it is essential to use multiple independent observables simultaneously to definitively understand the role of bursty star formation in high-$z$ galaxies' evolution \citep[see also e.g.,][]{Mirocha23,Endsley23b,Topping2024,Langeroodi24,Trussler24}.

Since our model assumes a redshift-independent star formation efficiency and still struggles to reproduce the extremely high redshift observations at $z>12$, there is a clear need for further investigation of the physics of the earliest galaxies to understand how they formed the first stellar populations.

\section*{}
We thank Andrea Ferrara, Anne Hutter, Julian Mu\~{n}oz and Stefania Salvadori for useful discussions, and Dan Stark for providing comments on a draft of the manuscript.
VG and CAM acknowledge support by the Carlsberg Foundation under grant CF22-1322. The Cosmic Dawn Center (DAWN) is funded by the Danish National Research Foundation under grant DNRF140.

\appendix

\section{The impact of cosmological assumptions on the UVLF} \label{sec:cosmoassumptions}
To make our predictions for the UVLFs, we had to make some assumptions regarding: (i) the cosmological parameters, (ii) the halo mass function, and (iii) the form used for the matter transfer function. Previous work modelling high-$z$ LFs have assumed a variety of choices for the above.
To assess the impact of these choices on our results, we have tested our UV-scatter model results varying these set of assumptions one at a time.

By changing the cosmological parameters from those of a flat $\Lambda$CDM with $\Omega_m=0.3$, $\Omega_\Lambda=0.7$ and $h=0.7$, to the \cite{Planck18} measurements through the cosmic microwave background (CMB) anisotropies ($\Omega_m=0.315$, $\Omega_\Lambda=0.685$ and $h=0.67$), we find a completely negligible change in the LF.

\begin{figure}[h!]
    \centering
    \includegraphics[width=0.75\textwidth]{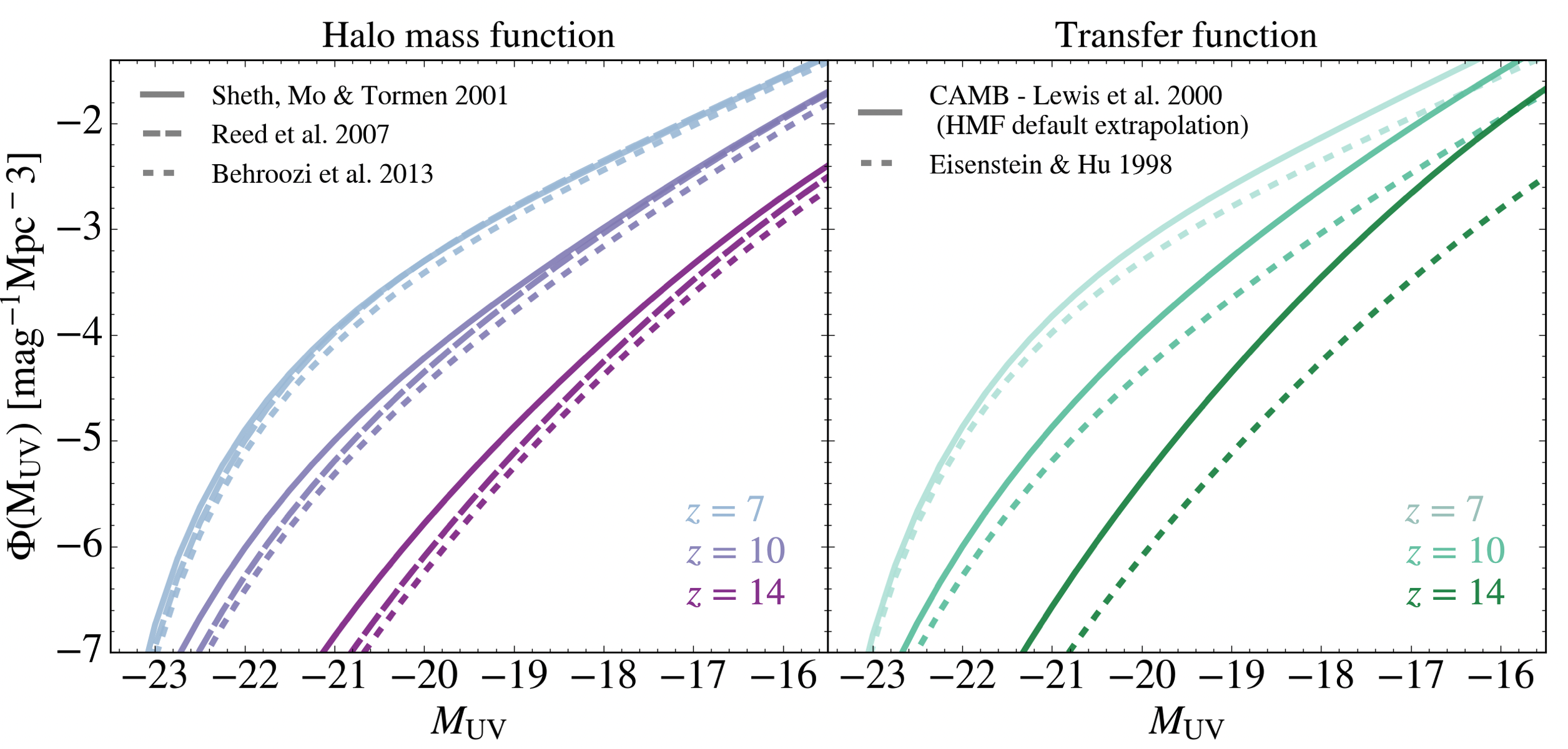}
    \caption{UVLF at $z=7,10,14$ for different halo mass functions (\textit{left}) and transfer functions (\textit{right}).
    \label{fig:hmf_transf}
    }
\end{figure}

For the HMF, we use the \code{HMF} python package by \cite{Murray13_hmf} and assumed a \cite{Reed07} function, which had been simulated specifically for halos at $z\sim10-30$. Using instead the functions from \cite{Sheth01} and \cite{Behroozi13} results in slightly higher and lower number densities respectively, with the maximum change produced in the LFs of $\sim0.5$~dex (see Figure~\ref{fig:hmf_transf}, left panel). We can conclude that the choice of the HMF produces a minor effect on the LFs, but that our overall conclusions do not depend on its choice.

Finally, we test the impact of the assumed linear transfer function by changing it from the \cite{EisensteinHu98} form to the one calculated with the Code for Anisotropies in the Microwave Background \citep[CAMB, ][]{Lewis00_camb}, both included in the \code{hmf} package. This time, the change in the LFs is significant, especially towards the faint end, with the LF being higher by up to $\sim 1$~dex when using CAMB. This could be linked to an excess of the power spectrum at very small scales in the CAMB function implementation, and/or possibly due to the default extrapolation of the CAMB transfer function at high wavenumber in the \code{HMF} code\footnote{\url{https://github.com/halomod/hmf/issues/90}}. We therefore caution that care must be taken when using the default CAMB option for the transfer function in \code{HMF} as it can lead to an overestimation of the LF at high-$z$.

\bibliography{refer,codes}
\bibliographystyle{aasjournal}

\end{document}

%% file: pacchetti.tex
\usepackage[english]{babel}
\usepackage{amsmath}
\usepackage{amssymb}
\usepackage{graphicx}
\usepackage{subfigure}
\usepackage[normalem]{ulem}
\usepackage{enumerate}
\usepackage{booktabs}

\usepackage{verbatim}

\usepackage{color}
\usepackage{multirow}
\usepackage{mathtools}
\usepackage{epstopdf}
\usepackage{tabularx}

\usepackage{url}
\usepackage{xcolor}

%% file: journals.tex
\def\apjl{ApJL}
\def\apjs{ApJS}
\def\aap{A\&A}

%% file: definitions.tex
\def\be{\begin{equation}} 
\def\ee{\end{equation}} 
\def\ba{\begin{eqnarray}} 
\def\ea{\end{eqnarray}}

\def\msun{{\Msun}}

\def\gsim{\lower.5ex\hbox{\gtsima}} 
\def\lsim{\lower.5ex\hbox{\ltsima}} \def\gtsima{$\; \buildrel > \over 
\sim \;$} \def\ltsima{$\; \buildrel < \over \sim \;$} \def\prosima{$\; 
\buildrel \propto \over \sim \;$} \def\gsim{\lower.5ex\hbox{\gtsima}} 
\def\lsim{\lower.5ex\hbox{\ltsima}} 
\def\simgt{\lower.5ex\hbox{\gtsima}} 
\def\simlt{\lower.5ex\hbox{\ltsima}} 
\def\simpr{\lower.5ex\hbox{\prosima}}   
  
 \def\gtsima{$\; \buildrel > \over \sim \;$} 
\def\ltsima{$\; \buildrel < \over \sim \;$} 
\def\gsim{\lower.5ex\hbox{\gtsima}} 
\def\lsim{\lower.5ex\hbox{\ltsima}} 
\def\simgt{\lower.5ex\hbox{\gtsima}} 
\def\simlt{\lower.5ex\hbox{\ltsima}} 
\def\simpr{\lower.5ex\hbox{\prosima}}

\def\msun{\,{\rm \Msun}}

\def\E3{{\cal E}_{\rm g}^{III}}

\def\r12{r_{1/2}} 
\def\x12{x_{1/2}} 
\def\v12{v_{1/2}}

\newcommand\code[1]{\textsc{\MakeLowercase{#1}}}

\def\nh2{n_{\rm H2}}
\def\fh2{f_{\rm H2}}

\def\angstrom{\textrm{A\kern -1.3ex\raisebox{0.6ex}{$^\circ$}}}

\def\msun{{\rm M}_{\odot}}

\def\msunyr{\msun\,{\rm yr}^{-1}}

\def\muv{M_{\rm UV}}
\def\siguv{\sigma_{\rm UV}}